\documentclass[12pt,apj]{emulateapj}
\slugcomment{}
\journalinfo{submitted to AJ August 15, 2006; accepted November 28, 2006}
\shorttitle{BALMER-LINE BAL QUASAR}
\shortauthors{Hall et al.}
\newcommand{\bb}{SDSS~J1259+0931}

\begin{document}
\title{A Balmer-Line Broad Absorption Line Quasar}
\author{Patrick B. Hall}
\affil{Department of Physics and Astronomy,
York University, 4700 Keele St., Toronto, Ontario M3J 1P3, Canada}

\begin{abstract}
I report the discovery of blueshifted broad absorption line (BAL) troughs 
in at least six transitions of the Balmer series of hydrogen (H$\beta$ to H9) 
and in \ion{Ca}{2}, \ion{Mg}{2} and excited \ion{Fe}{2} in the quasar SDSS 
J125942.80+121312.6.  This is only the fourth active galactic nucleus known to 
exhibit Balmer absorption, all four in conjunction with low-ionization BAL 
systems containing excited \ion{Fe}{2}.  The substantial population in the
$n=2$ shell of \ion{H}{1} in this quasar's absorber likely arises from
Ly$\alpha$ trapping.  In an absorber sufficiently optically thick to
show Balmer absorption, soft X-rays from the quasar penetrate to large
$\tau_{Ly\alpha}$ and ionize \ion{H}{1}.  Recombination then creates Ly$\alpha$
photons that increase the $n=2$ population by a factor $\tau_{Ly\alpha}$ since
they require $\simeq\tau_{Ly\alpha}$ scatterings to diffuse out of the absorber.
Observing Ly$\alpha$ trapping in a quasar absorber requires a large but
Compton-thin column of gas along our line of sight which includes substantial
\ion{H}{1} but not too much dust.
Presumably the rarity of Balmer-line BAL troughs 
reflects the rarity of such conditions in quasar absorbers.
\end{abstract}

\section{Background} \label{intro}

Broad absorption line (BAL) active galactic nuclei (AGN) 
show absorption from gas at blueshifted velocities up to 0.2$c$.
They are seen in about one in four of the luminous AGN known as quasars,
and while most BAL AGN only show high-ionization absorption 
(e.g., \ion{N}{5}, \ion{C}{4}),
$\sim$10\% of BAL AGN also exhibit low-ionization absorption 
(e.g., \ion{Al}{3}, \ion{Mg}{2}), 
including $\sim$3\% which in addition show excited \ion{Fe}{2} absorption
(\nocite{trump06}{Trump} {et~al.} 2006, and references therein).
If all quasars have BAL troughs, these percentages represent the fractions of
unobscured lines of sight toward quasars along which these three types of
troughs exist.  If all quasars go through a BAL phase, they represent the 
fractional amount of time each type exists.  
Reality is likely somewhere in between.
On the one hand, BAL and non-BAL quasar spectra are very similar, arguing that
they are drawn from a single parent population \nocite{wea91,sdssbal}({Weymann} {et~al.} 1991; {Reichard} {et~al.} 2003).
On the other hand, 
the likelihood of observing a low-ionization BAL trough
increases
during a transition phase between ultraluminous infrared galaxies and quasars
\nocite{cs01}({Canalizo} \& {Stockton} 2001), although the origin of the BAL troughs in such objects is
an open question; a disk wind origin for them remains possible.

The physical parameters of BAL systems need to be constrained to understand
their origin(s) and contribution to AGN feedback effects on galaxy formation
and evolution.  However, saturation and blending effects mean that the 
physical parameters of typical BAL systems cannot be easily inferred.  
Instead, it is often the more unusual systems that can provide such 
constraints.  Currently the rarest known absorption in BAL AGN is Balmer-line
absorption.  It has previously been reported only in NGC~4151 (\nocite{hut02}{Hutchings} {et~al.} 2002,
and references therein) and SDSS~J083942.11+380526.3 \nocite{aoki06}({Aoki} {et~al.} 2006).
Weak absorption similar to that found in those objects
has also recently been seen in H$\beta$ and H$\gamma$ 
in the quasar FBQS J2107$-$0620 through high-resolution spectroscopy 
(Hutsemekers et al., in preparation).
Here I report a quasar with very strong Balmer-line absorption 
found in the Sloan Digital Sky Survey Data Release Five
\nocite{dr5}({Adelman-McCarthy} {et~al.} 2007).
These two new discoveries confirm the observation of \nocite{aoki06}{Aoki} {et~al.} (2006) 
that Balmer-line BAL troughs are found in iron low-ionization BAL quasars 
with relatively strong [\ion{O}{3}] emission.

\section{Observations} \label{obs}

The Sloan Digital Sky Survey \nocite{yor00}(SDSS; {York} {et~al.} 2000) is using a drift-scanning
imaging camera \nocite{gun98}({Gunn} {et~al.} 1998) on a 2.5-m telescope \nocite{gun06}({Gunn} {et~al.} 2006) to image
10$^4$\,deg$^2$ of sky on the SDSS $ugriz$ AB magnitude system
\nocite{fuk96,sdss82,sdss105,sdss153,ive04}({Fukugita} {et~al.} 1996; {Hogg} {et~al.} 2001; {Smith} {et~al.} 2002; {Pier} {et~al.} 2003; {Ivezi{\'c}} {et~al.} 2004).
Two multi-fiber, double spectrographs are being used to obtain $R\sim1900$
spectra for $\sim$10$^6$ galaxies to $r=17.8$ and $\sim$10$^5$ quasars to 
$i=19.1$ ($i=20.2$ for $z>3$ candidates).  As discussed in \nocite{sdssqtarget}{Richards} {et~al.} (2002),
quasar candidates are targeted for spectroscopy because their colors differ
from the colors of the stellar locus or because they are unresolved objects 
with radio emission detected by the FIRST survey \nocite{bwh95}({Becker}, {White}, \& {Helfand} 1995).

SDSS J125942.80+121312.6 (hereafter SDSS J1259+1213) was observed on Modified
Julian Date 53473 with SDSS plate \#1695 and fiber \#75.
It has a Galactic extinction corrected magnitude $i=18.15\pm0.02$
and was targeted both via its colors and its radio emission.
Its FIRST peak flux of $2.00\pm 0.14$ mJy/beam puts it right on the
border between radio-loud and radio-quiet, with $R_i=1.00$ \nocite{sdss1st}({Ivezi{\' c}} {et~al.} 2002).  

Figure 1 shows that
the quasar has a blue continuum with P Cygni-like emission and absorption
in \ion{Mg}{2}, excited-state \ion{Fe}{2}
and the Balmer series (H$\beta$ through H9, and possibly H10 and H11).
The redshift of the quasar, measured
from [\ion{O}{2}] and consistent with that of
[\ion{O}{3}] $\lambda$5007, is $z=0.7517\pm0.0001$, yielding $M_i=-24.88$
($\lambda_iL_{\lambda_i}=2\times 10^{45}$ ergs s$^{-1}$).  
The broad \ion{Mg}{2} and H$\beta$ emission lines peak at $z=0.748\pm0.001$,
slightly blueshifted from the narrow-line redshift.  
The deepest Balmer absorption is blueshifted further, to $z=0.7345\pm0.0004$,
which is $2960\pm 70$\,km\,s$^{-1}$\ from the narrow-line redshift.  
The continuum drop shortward of $\sim$2600\,\AA\ rest frame is due to 
partial covering of the quasar by overlapping absorption from 
multiple \ion{Fe}{2} transitions (e.g., \nocite{sdss123}{Hall} {et~al.} 2002).
The smaller absorption depths for neutral \ion{H}{1} as compared to
low-ionization \ion{Fe}{2} and \ion{Mg}{2} are consistent with an 
increasing covering factor with ionization level, often seen in BAL quasars.

Emission and absorption in higher-excitation transitions of
\ion{Fe}{2}, \ion{Ti}{2} and \ion{Cr}{2}
\nocite{vea06}({V{\'e}ron-Cetty} {et~al.} 2006) can
plausibly explain the complex continuum between \ion{Mg}{2} and [\ion{O}{2}].
(Similar absorption at 2800\,\AA\,$<$\,$\lambda$\,$<$\,3500\,\AA\ is seen in
the low-ionization BAL quasar SDSS J112526.13+002901.3; \nocite{sdss123}{Hall} {et~al.} 2002.)
For example, the lower term of \ion{Fe}{2} $d^2D^1-x^2P^o$,
tentatively identified in Figure 1, is at 5.91 eV above ground.
There is no sign of metastable \ion{He}{1} absorption,
but the limits are not strong since those transitions
suffer from confusion with absorption from \ion{Fe}{2} and H8.
Lastly, the [\ion{O}{3}] emission-line profile may have a complex blueshifted
component, or it may just be confused with the strong \ion{Fe}{2} emission 
in that region of the spectrum.\footnote{Note that the standard optical 
\ion{Fe}{2} emission template derived from the spectrum of I~Zw~1 \nocite{bg92}({Boroson} \& {Green} 1992)
is not a particularly good match to the \ion{Fe}{2} emission in \bb.}

Balmer line absorption is often seen in galaxy spectra, 
but this object cannot be explained as an unusual galaxy.
First, the emission and broad absorption in ultraviolet \ion{Fe}{2} and
\ion{Mg}{2} transitions is consistent only with a BAL quasar.
Second, when post-starburst galaxies exhibit strong Balmer absorption,
it lies at the same redshift as the [\ion{O}{2}] and [\ion{O}{3}] emission
and is accompanied by a strong Balmer break lacking in this object.
Third, when low-ionization outflows are seen in star-forming galaxies, 
they do not occur at velocities as high as 3000~km\,s$^{-1}$ 
unless they are clearly BAL troughs associated with an AGN \nocite{rvs05}({Rupke}, {Veilleux}, \& {Sanders} 2005).

\section{Interpretation} \label{interp}

The Balmer absorption troughs in this object have widths which are uniform 
within the measurement uncertainties: ${\rm FWZI}=2000\pm 200$~km\,s$^{-1}$.
To estimate the depths of the troughs, the continuum was fitted with a
fourth-order polynomial in $F_{\lambda}$ vs. $\log\lambda$ to approximate a
power-law with possible slight dust reddening (see Figure 1).
The maximum single-pixel depth of each trough was measured
from the normalized SDSS spectrum after smoothing by a 7-pixel boxcar.
The depths of the troughs (Table 1) decrease slowly
as the upper term of the transition increases.
The decline is about a factor of two from H$\beta$ to H9.
The absorption therefore must be saturated, since the transition oscillator
strengths decline by a factor of {\em twenty-two} from H$\beta$ to H9.

The percentage absorption depth in a trough $l$ is $D_l=C(1-e^{-\tau_l})$ for an
absorber with optical depth $\tau_l$ and percentage covering factor $C$ in the 
relevant ion (e.g., Eq.~1 of \nocite{sb2}{Hall} {et~al.} 2003).
By assuming $C$ and $\tau_{H\beta}$, 
$D_l$ for every Balmer trough can be calculated since their relative $\tau$ are 
determined by their known oscillator strengths.
Acceptable matches to the maximum absorption depths of the Balmer lines
were found with $\tau_{H\beta}=19.5\pm2.5$ and $C=29\mp1\%$ at the velocity of
maximum depth (Table 1); note that larger $\tau$ requires smaller $C$ for an
equally good fit. The additional depth in the H$\beta$ trough is assumed to come
from coverage of the H$\beta$ emission line.\footnote{If the total unabsorbed
flux has continuum and line components, $F=F_C+F_L$, then the flux removed by
very high optical depth absorption ($\tau\rightarrow\infty$) is 
$A=C_CF_C+C_LF_L$, 
allowing for different covering factors $C$ of each component.  
Since the normalization only used the continuum, the depths in Table 1 are
$D=A/F_C=C_C+C_LF_L/F_C$.  Thus, $D>C_C$ is possible when the line flux $F_L$ 
is substantial, as is the case for H$\beta$ in this object.
}
This approximate $\tau_{H\beta}$ assumes the Balmer-line troughs are resolved
and takes into account neither possible Balmer emission from the absorber 
itself (as in the resonance scattering model of \nocite{bra02}{Branch} {et~al.} 2002; 
see also \nocite{cbbl04}{Casebeer} {et~al.} 2004) nor possible different
\ion{H}{1} covering factors for the continuum source, Balmer emission lines 
and \ion{Fe}{2} emission-line blends.
A high-resolution spectrum is needed to properly model all those parameters
and determine how greatly they affect results inferred from an SDSS-resolution
spectrum.

Absorption from the $n=2$ shell of hydrogen requires a substantial
population in that shell.   One obvious population mechanism is
collisional excitation into $n=2$ due to a high density.
The relevant critical density 
is $n_{crit} = 8.7\times 10^{16} \tau_{Ly\alpha}^{-1}\ {\rm cm}^{-3}$ 
at $T$=10,000~K \nocite{of06}({Osterbrock} \& {Ferland} 2006).  
It is possible to estimate $\tau_{Ly\alpha}$ using 
our estimate for $\tau_{H\beta}$ and 
the relationship\footnote{This relationship
comes from the fact that for any transition originating in shell $k$,
$n_k\propto N_H(n=k)\propto\tau_{k}\Delta v/\lambda_{k}f_{k}$ (e.g., Eq.~12 of
\nocite{sb2}{Hall} {et~al.} 2003). Thus, $n_1/n_2=\tau_1\lambda_2f_2/\tau_2\lambda_1f_1$, assuming
an absorber of fixed width $\Delta v$ with all $\tau$ independent of velocity.
}
$\tau_{Ly\alpha}=\frac{7n_1}{8n_2}\tau_{H\beta}$, 
where the ratio of the populations of the $n=1$ and $n=2$ shells is
$\frac{n_1}{n_2}\simeq\frac{1}{4}\exp(10.2\ {\rm eV}/kT)$.
For $T$=10,000~K, $\frac{n_1}{n_2}=3.50\times 10^4$, and so
$\tau_{Ly\alpha}\simeq(6.6\pm0.8)\times 10^{5}$ when $\tau_{H\beta}=19.5\pm2.5$.
This would imply $n_e\simeq (1.3\pm0.2) \times 10^{11}\ {\rm cm}^{-3}$
if collisional excitation was the only mechanism populating $n=2$.

However, at such large $\tau_{Ly\alpha}$ there is another effect which 
populates the $n=2$ shell: Ly$\alpha$ trapping \nocite{fn79}({Ferland} \& {Netzer} 1979).
As shown in the previous paragraph,
an absorber with Balmer optical depths as large as observed
here must have a very large Ly$\alpha$ optical depth.  In such an absorber,
soft X-rays from the quasar will ionize \ion{H}{1} at large $\tau_{Ly\alpha}$ 
and create Ly$\alpha$ photons at those depths via recombination.  
Every Ly$\alpha$ photon so created will be re-absorbed approximately 
$\tau_{Ly\alpha}$ times before it escapes, and so the $n=2$ population for a 
given density and temperature will be increased by a factor of 
$\tau_{Ly\alpha}$.

This increase changes the value of $\tau_{Ly\alpha}$ inferred for a given 
$\tau_{H\beta}$, and thus the \ion{H}{1} column density needed to explain 
a given $\tau_{H\beta}$.
When Ly$\alpha$ trapping is occurring, $\tau_{Ly\alpha}^{trap} \simeq [\tau_{H\beta}\frac{7}{32}\exp(10.2\ {\rm eV}/kT)]^{1/2}$.
For our observed $\tau_{H\beta}$ and $T=7500$\,K, appropriate for partially
ionized gas illuminated by a quasar \nocite{of06}(Figure 14.5 of {Osterbrock} \& {Ferland} 2006), $\tau_{Ly\alpha}=5520$ and
$\frac{n_1}{n_2}\simeq\frac{1}{4}(\tau_{Ly\alpha}^{trap})^{-1}\exp(10.2\ {\rm eV}/kT)=323$.
The H$\beta$ optical depth yields an estimate of the column density of neutral
hydrogen in the $n=2$ shell (Eq.~12 of \nocite{sb2}{Hall} {et~al.} 2003),
assuming a uniform optical depth over the entire 2000 km s$^{-1}$ trough:
$N_{\rm HI}(n=2) \simeq (2.5\pm0.3) \times 10^{16}\ {\rm cm}^{-2}$.
Combining the estimates for $N_{\rm HI}(n=2)$ and $\frac{n_1}{n_2}$ yields 
$N_{\rm HI}\simeq8.1\times10^{18}\ {\rm cm}^{-2}$ in the absorber.
As is usual for BAL AGN,
$N_{\rm H} = N_{\rm HI} + N_{\rm HII}$ is likely to be much higher.

Ly$\alpha$ trapping eliminates the need for an absorber with a density 
above the critical density needed
for significant collsional excitation.  
In fact, $n_{crit}$ becomes an {\em upper} limit on the
density in the absorber
in order to avoid {\em overpopulating} the $n=2$ shell 
through collisions as well as Ly$\alpha$ trapping.  Since that would produce
Balmer-line optical depths greater than are observed, the density 
in an absorber at $T=7500$\,K must be $n_e < 1.6 \times 10^{13}\ {\rm cm}^{-3} = 8.7\times 10^{16}(\tau_{Ly\alpha}^{trap})^{-1}\ {\rm cm}^{-3}$.

For Ly$\alpha$ trapping to occur, the absorber 
must retain a relatively large neutral hydrogen column
despite exposure to the quasar's ionizing continuum.
Quasar absorbers with blueshifted neutral hydrogen
columns that large located relatively close to the quasar 
are rare along our sightlines to bright quasars,
either because such columns cover a very small total solid angle
or are very short-lived phenomena.
However, Ly$\alpha$ trapping is expected to occur in
quasar low-ionization broad emission line regions.  \bb\ may be a
case where the continuum source is partially viewed through such gas.

Finally, it should be mentioned that an absorber with a high temperature
rather than a high $\tau_{Ly\alpha}^{trap}$ could in principle
produce the observed Balmer absorption.  For example, $T=85,200$~K
yields equal populations in the $n=1$ and $n=2$ shells of \ion{H}{1}.  
However, this explanation is unlikely 
because gas at such temperatures is unstable to rapid cooling \nocite{kro99}({Krolik} 1999).

\section{Conclusions} \label{concl}

The quasar \bb\ has
the strongest Balmer-line BAL troughs discovered to date.  
The required population of the $n=2$ shell of \ion{H}{1} most likely
arises from Ly$\alpha$ trapping in an absorber exposed to the quasar's
ionizing continuum but still containing a high column of \ion{H}{1}
($N_{\rm HI}\simeq8.1\times10^{18}\ {\rm cm}^{-2}$).  In addition, 
the detection of this quasar in the rest-frame optical and ultraviolet
means the absorber is neither Compton-thick nor particularly dust-rich.
The rarity of such Balmer-line BAL troughs 
(found in $\lesssim$1\% of BAL quasars)
indicates they have a very small global covering factor or are very transient.
However, their incidence may be underestimated because for high-redshift BAL
AGN the Balmer lines are redshifted beyond the typical spectral coverage of a
discovery spectrum.  Even when that is not the case, weak Balmer absorption
can easily be missed (e.g., the absorption in FBQS J2107$-$0620 was discovered
only through high-resolution spectroscopic followup).

To improve our understanding of quasar winds requires observations of 
specific objects for which multiple physical parameters in the absorbers can
be constrained.  Thus, follow-up observations of \bb\ would be quite valuable.
X-ray observations can constrain the total absorbing column.
Echelle spectroscopy can put a lower limit on the ionized hydrogen 
column by searching for absorption from metastable \ion{He}{1}, 
which is populated by recombination from \ion{He}{2}.  Such spectroscopy would
also reveal any variability or structure in the Balmer-line troughs and enable
detailed modelling of the Balmer and excited \ion{Fe}{2} emission and
absorption to constrain the absorber's density and ionization state
and thus its distance from the ionizing source.
A near-IR spectroscopic search for Paschen absorption can constrain the
population in the $n=3$ shell of \ion{H}{1}.
Near-IR spectroscopy of H$\alpha$ can help constrain the contribution of any
`fill-in' Balmer emission from the absorber itself.  Such emission should
be $\sim$4 times stronger in H$\alpha$ than in H$\beta$ given the Balmer and
inferred Lyman optical depths of the absorber
\nocite{cm69}({Cox} \& {Mathews} 1969).
\bb\ is detected by 2MASS with $J=17.00\pm0.16$ and $K_s=15.32\pm0.15$,
placing moderate resolution near-IR spectroscopy within the reach of 8m-class
telescopes.

\acknowledgments
I thank N. Murray, D. Hutsemekers, and the referee.
P. B. H. is supported by NSERC.
    Funding for the SDSS and SDSS-II has been provided by the Alfred P. Sloan Foundation, the Participating Institutions, the National Science Foundation, the U.S. Department of Energy, the National Aeronautics and Space Administration, the Japanese Monbukagakusho, the Max Planck Society, and the Higher Education Funding Council for England. The SDSS Web Site is http://www.sdss.org/.
    The SDSS is managed by the Astrophysical Research Consortium for the Participating Institutions. The Participating Institutions are the American Museum of Natural History, Astrophysical Institute Potsdam, University of Basel, Cambridge University, Case Western Reserve University, University of Chicago, Drexel University, Fermilab, the Institute for Advanced Study, the Japan Participation Group, Johns Hopkins University, the Joint Institute for Nuclear Astrophysics, the Kavli Institute for Particle Astrophysics and Cosmology, the Korean Scientist Group, the Chinese Academy of Sciences (LAMOST), Los Alamos National Laboratory, the Max-Planck-Institute for Astronomy (MPIA), the Max-Planck-Institute for Astrophysics (MPA), New Mexico State University, Ohio State University, University of Pittsburgh, University of Portsmouth, Princeton University, the United States Naval Observatory, and the University of Washington.
This research has also made use of
the Atomic Line List v2.04 at \url{http://www.pa.uky.edu/$\sim$peter/atomic/}.

\begin{deluxetable}{lccccccccc}
\tabletypesize{\small}
\tablecaption{Absorption Parameters in SDSS J125942.80+121312.6\label{t_1}}
\tablewidth{500pt}
\tablehead{
\colhead{} &
\colhead{H$\beta$} &
\colhead{H$\gamma$} &
\colhead{H$\delta$} &
\colhead{H7\tablenotemark{a}} &
\colhead{H8} &
\colhead{H9} &
\colhead{\ion{Ca}{2},K} &
\colhead{\ion{Mg}{2}} &
\colhead{\ion{Fe}{2}}
}
\startdata
Observed Depth\tablenotemark{b}&	36\% &	29\% &	27\% &	23\% &	18\% &	17\% &	7\% &	58\% &	$\sim$50\% \\
Model Depth\tablenotemark{b}&		29.0\%&	29.0\%&	27.6\%&	23.7\%&	18.9\%&	14.6\%&	\nodata & \nodata & \nodata \\
$f_{ij}$\tablenotemark{c}& 0.119& 0.0446& 0.0221& 0.0127& 0.00803& 0.00543& \nodata& 0.943& \nodata \\
$\lambda$& 4862.683& 4341.684& 4102.892& 3971.195& 3890.151& 3836.472& 3934.777& $\lambda\lambda$2798.75& $<$2632.106\\
\enddata
\tablenotetext{a}{H7 is blended with \ion{Ca}{2},H $\lambda$3969.591, but that
does not affect measurement of its observed depth.}
\tablenotetext{b}{Uncertainties on the observed depths are 2-3\% (1$\sigma$).
The model depths are given for $\tau_{H\beta}=19.5$ and $C=29\%$.
For further details on the observed and model depths, see \S\ \ref{interp}.
}
\tablenotetext{c}{The $f_{ij}$ values are the oscillator strengths,
and can be used along with the wavelengths to calculate
predicted relative strengths of the Balmer absorption troughs.
For reference, $f_{ij}=0.640$ for H$\alpha$ $\lambda$6564.61
and $f_{ij}=0.416$ for Ly$\alpha$ $\lambda$1216.6701.
The $f_{ij}$ value for \ion{Mg}{2} is the sum for both members of the doublet.}
\end{deluxetable}

\begin{figure}
\epsscale{1.1}
\plotone{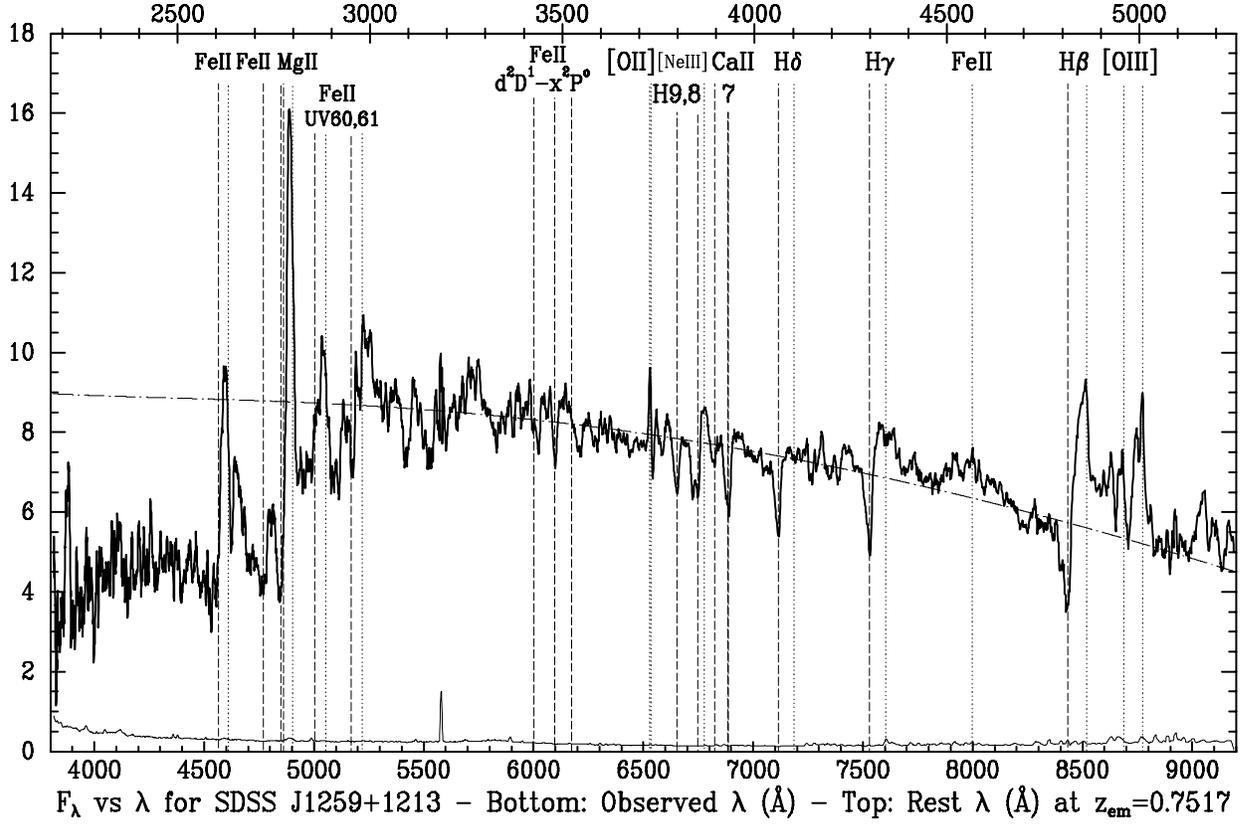}
\caption{Full SDSS spectrum of
SDSS J125942.80+121312.6, smoothed by a 7 pixel boxcar, with $F_\lambda$ in
units of 10$^{-17}$ erg s$^{-1}$ cm$^{-2}$ \AA$^{-1}$ and $\lambda$ in \AA.
Uncertainties in the smoothed flux are shown by the thinner line
along the bottom of the plot.
Observed wavelengths are shown on the bottom axis, while rest frame wavelengths
at the narrow-line emission redshift of $z=0.7517$ are shown along the top axis.
Dotted lines show the expected location of emission lines at the emission
redshift, while dashed lines show absorption at the peak absorption redshift 
of $z=0.7345$.  The dot-dashed line shows the fit to the effective continuum.
\label{f_1}}
\end{figure}

\end{document}